\documentclass[twocolumn,aps,prd,preprintnumbers,superscriptaddress,nofootinbib,10pt]{revtex4-1}
\usepackage{epsfig}
\usepackage{graphicx}
\usepackage{psfrag}
\usepackage{amsmath,amssymb}
\usepackage{colordvi}
\usepackage{amsfonts}
\usepackage{enumerate}
\usepackage{slashed}
\usepackage{color}
\usepackage{xcolor}

\newcommand{\bs}[1]{\boldsymbol{#1}}

\begin{document}
\title{Gluon Gravitational Form Factors at Large Momentum Transfer }

\author{Xuan-Bo Tong}
\affiliation{ CAS Key Laboratory of Theoretical Physics, Institute of Theoretical Physics, Chinese Academy of Sciences, Beijing 100190, China}
\affiliation{School of Physical Sciences,University of Chinese Academy of Sciences,Beijing 100049,China}

\affiliation{Nuclear Science Division, Lawrence Berkeley National
Laboratory, Berkeley, CA 94720, USA}

\author{Jian-Ping Ma}
\affiliation{ CAS Key Laboratory of Theoretical Physics, Institute of Theoretical Physics, Chinese Academy of Sciences, Beijing 100190, China}

\author{Feng Yuan}
\affiliation{Nuclear Science Division, Lawrence Berkeley National
Laboratory, Berkeley, CA 94720, USA}

\begin{abstract}
We perform a perturbative QCD analysis of the gluonic gravitational form factors (GFFs) of the proton and pion at large momentum transfer. We derive the explicit factorization formula of the GFFs in terms of the distribution amplitudes of hadrons. At the leading power, we find that $A_g^\pi(t)=C_g^\pi(t)\sim 1/(-t)$ for pion, $A_g^p(t)\sim 1/(-t)^2$ and $C_g^p(t)\sim \ln^2(-t/\Lambda^2)/(-t)^3$ for proton, respectively, where $t$ is the momentum transfer and $\Lambda$ a non-perturbative scale to regulate the endpoint singularity in $C_g^p$ calculation. Our results provide a unique perspective of the momentum dependence of the GFFs and will help to improve our understanding of the internal pressure distributions of hadrons.  \end{abstract}
\maketitle

\section{Introduction}

The gravitational form factors (GFFs) are the fundamental ingredients to probe the internal structure of hadrons.  As the matrix elements of the energy-momentum tensor~(EMT)~\cite{Kobzarev:1962wt,Pagels:1966zza,Ji:1994av,Ji:1995sv,Ji:1996ek,Ji:1996nm}, they provide important information on the hadron's mass and spin~\cite{Pagels:1966zza,Ji:1994av,Ji:1995sv,Ji:1996ek,Ji:1996nm,Metz:2020vxd,Jaffe:1989jz,Filippone:2001ux,Bass:2004xa,Aidala:2012mv,Leader:2013jra,Ji:2016djn,Deur:2018roz,Ji:2020ena}, and the mechanical property~\cite{Polyakov:2002yz,Polyakov:2018zvc,Burkert:2018bqq,Shanahan:2018nnv,Kumericki:2019ddg}. In experiments, the GFFs can be constrained from the generalized parton distributions (GPD)~\cite{Ji:1996ek,Ji:1996nm,Mueller:1998fv,Diehl:2003ny,Belitsky:2005qn} which are measured in the  hard exclusive processes like deeply virtual Compton scattering~\cite{Ji:1996ek,Ji:1996nm,Radyushkin:1996nd,dHose:2016mda,Kumericki:2016ehc} and deeply virtual meson production~\cite{Collins:1996fb,Mankiewicz:1997bk,Favart:2015umi}. 

Recently, a glimpse of the quark GFFs and its interpretation as a pressure distribution inside the proton has been reported in Ref.~\cite{Burkert:2018bqq}. The lattice QCD has also been applied to compute the GFFs for the quarks and gluons~\cite{Gockeler:2003jfa,Gockeler:2005cj,Brommel:2005ee,Brommel:2007zz,Hagler:2003jd,Hagler:2007xi,Hagler:2009ni,Gockeler:2006zu,Alexandrou:2011nr,Shanahan:2018pib} and deep insight has been obtained from these studies~\cite{Shanahan:2018nnv}. All these developments have attracted great attention in the hadron physics community and it is expected that future measurements at both JLab 12 GeV~\cite{Dudek:2012vr} and the Electron-Ion collider~\cite{Accardi:2012qut,Boer:2011fh} will provide more important constraints on the quark/gluon GFFs of the hadrons.

In this paper, we will investigate the GFFs at large momentum transfer, focusing on the gluonic contributions. This will provide a unique perspective of their behaviors and improve the parameterizations in the wide range of kinematics. At large momentum transfer, the form factors can be calculated from perturbative QCD ~\cite{Lepage:1979za,Brodsky:1981kj,Efremov:1979qk,Chernyak:1977as,Chernyak:1980dj,Chernyak:1983ej,Belitsky:2002kj}. Previously, a power counting method~\cite{Brodsky:1973kr,Matveev:1973ra,Ji:2003yj} was applied to estimate the power behaviors for the quark GFFs~\cite{Tanaka:2018wea}. The power behavior arguments have also played important roles in the phenomenology studies~\cite{Burkert:2018bqq,Shanahan:2018pib,Frankfurt:2002ka}. The factorization formalism for the GFFs follows that developed in the literature for the hard exclusive processes at large momentum transfer and the final results depend on the gauge invariant distribution amplitudes of hadrons~\cite{Burkardt:2002uc,Ji:2002xn,Ji:2003fw,Hoodbhoy:2003uu,Braun:1999te,Braun:2000kw}.

Meanwhile, the gluon GFFs of nucleon play important roles in the near threshold heavy quarkonium photo-productions. These processes have gained quite an interest in recent years, because they promise to measure the proton mass decomposition~\cite{Kharzeev:1998bz,Brodsky:2000zc,Gryniuk:2016mpk,Hatta:2018ina,Hatta:2019lxo,Boussarie:2020vmu,Mamo:2019mka,Gryniuk:2020mlh,Wang:2019mza,Zeng:2020coc,Du:2020bqj}. In the near threshold region, the momentum transfer from the nucleon target is relatively large, ($-t\sim 2{\rm GeV}^2$ and $10{\rm GeV}^2$ for $J/\psi$ and $\Upsilon$, respectively). Therefore, our results for the gluon GFFs at large momentum transfer shall make a valuable contribution to understanding the $t$-dependence in these processes. 

The gravitational form factors of the hadrons are the transition matrix elements of the energy momentum tensor. The gluon sector reads, 
\begin{align}
T_g^{\mu\nu}=G^{a\mu  \alpha} G^{a \nu}_{\alpha}+\frac{1}{4} g^{\mu \nu} G^a_{  \alpha \beta } G^{a\alpha \beta},\label{emt-gluon}
\end{align}
where $G^a_{\mu\nu}=\partial_\mu A^a_\nu-\partial_\nu A^a_\mu-g_s f^{abc}A^b_\mu A^c_\nu$ is the strength tensor of the gluon  field $A^a_\mu$. 
For the proton, the GFFs are parametrized as~\cite{Ji:1996ek,Ji:1996nm},
\begin{align}
 \langle P',s' | T_g^{\mu\nu}|P,s\rangle& =
\bar U_{s'}(P')  \left[ A_g(t)\gamma^{\{\mu } \bar P^{\nu\}}\right.\nonumber\\
&\left.+ C_g(t)\frac{\Delta^{\mu } \Delta^{\nu}-g^{\mu \nu}\Delta^2}{M} +\cdots 
\right] U_s(P) ,
 \label{EMT:proton}
 \end{align}
where $P$ and $P'$ are the initial and final state hadron momentum, respectively, $\Delta=P'-P$ is the momentum transfer and $t=\Delta^2$, $\bar P=(P+P')/2$ the average momentum, $a_{\{\mu}b_{\nu \}}=(a_\mu b_\nu +a_\nu b_\mu)/2$. $U_s(P)$ is the spinor of the nucleon with the spin $s$ and mass $M$, which is nomarlized as $\bar U_{s}(P) U_s(P)=2 M$. Here, we follow the notations in Refs.~\cite{Ji:1996ek,Ji:1996nm}, where $C$ form factor has also been referred as $D$ or $d_1$ form factor in Refs.~\cite{Polyakov:2002yz,Polyakov:2018zvc,Burkert:2018bqq,Shanahan:2018nnv,Kumericki:2019ddg} with different normalization: $D(t) = 4/5 d_1(t) = 4 C(t)$. 
In addition, we only keep the $A$ and $C$ form factors in the above equation for simplicity. 

All the gluon form factors depend on the renormalization scale, since the gluon piece of  EMT is not conserved individually and only the total GFFs are renormalization independent.  Generally, the $A$-form factors describe the distributions of the quark or gluon momentum inside the hadron, whereas the $C$-form factors characterize the mechanical properties. % in the hadron while the $\bar C $-form factors have intimate relationship with  the trace anomaly. In this paper, we will concentrate on the GFFs from the traceless components of EMT in the gluon sector.

In the following, we first show the derivations of the gluon GFFs of pion, where we compute both $A$ and $C $ form factors. Different from previous analysis, we find that both form factors scale as $1/(-t)$ at large momentum transfer. Then, we derive the gluon GFFs of nucleon. Different from the pion case, the nucleon's $C$ form factor is power suppressed respect to the $A$ form factor. The method developed in these calculations can be extended to all other form factors.

\section{Gravitational Form Factor for Pion}

\begin{figure}[tpb]
\includegraphics[width=0.5\columnwidth]{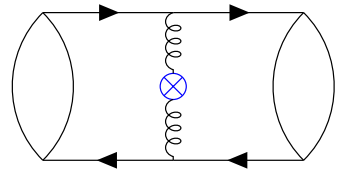}
    \caption{Leading order diagram contribution to the gluonic gravitational form factor of pion at large momentum transfer, where the incoming and out going hadron states have momenta $P$ and $P'$. The cross symbol in the middle of the diagram represents the operator of the gluonic component in the energy-momentum tensor of Eq.~(\ref{emt-gluon}).}
    \label{fig:pion}
\end{figure}

We start our analysis with the pion GFFs~\cite{Polyakov:2002yz,Polyakov:2018zvc,Shanahan:2018nnv},
\begin{align}
&\langle P' | T_g^{\mu\nu}|P\rangle
  =2 \bar  P^\mu  \bar  P^\nu A_g^{\pi}(t)\nonumber\\
 &~~~~~~~+\frac{1}{2} ( \Delta^\mu \Delta^\nu-g^{\mu\nu}\Delta^2) C_g^{\pi}(t) +2m^2 g^{\mu\nu }\overline{C}_g^\pi(t)\ , \label{EMT:pion}
 \end{align} 
where $m$ represents the pion mass. As shown in Fig.~\ref{fig:pion}, there is one diagram that contributes at the leading order of perturbation theory. The circle cross in the diagram denotes the local operator of the gluon EMT in Eq. (\ref{EMT:pion}).

Considering the leading asymptotic behaviour of large $-t$, the light-cone Fock state expansion of the pion have been performed with only minimal numbers of parton. The gluon EMT operator transport the two hard gluon exchanges between the quark line and generate the hard part of the GFFs.  Compared to the hard scale $t$, one can neglect the transverse momenta of partons in the hard part, since they are expected to be on the order of $\Lambda_{\text{QCD}}$. Integrating out the $k_\perp$ in the pion wave function, we obtain the disribution amplitude $\phi(x)=\int \frac{d^2 k_\perp}{(2\pi)^3} \psi(x,k_\perp)$~\cite{Burkardt:2002uc}. This finally leads to a factorization formula for the GFFs of the pion at large $t$: 
\begin{align}
&A^{\pi}_g(t,\mu)= \int d x_1 d y_1 \phi^*(y_1,\mu)\phi(x_1,\mu) {\cal A}^\pi_g(x_1,y_1,t,\mu)
,
\notag \\ &
C^{\pi}_g(t,\mu)=\int d x_1 d y_1 \phi^*(y_1,\mu)\phi(x_1,\mu) {\cal C}^\pi_g(x_1,y_1,t,\mu),
\label{eq:pionfac}
\end{align}
where $ {\cal A}^\pi_g$ or ${\cal C}^\pi_g$ is the perturbative calculable hard part of the GFFs. Before we present a detailed result for the hard part, a power counting analysis can be derived~\cite{Brodsky:1973kr,Matveev:1973ra}. The diagram of Fig.~\ref{fig:pion} is very similar to that for the electromagnetic form factor calculation at large momentum~\cite{Brodsky:1981kj}. Therefore, we can apply the same power counting and deduce that they should scale as $1/(-t)$ at large $-t$. Of course, we have to make sure that they do contribute to nonzero $A_g$ and $C_g$. 

Carrying out the calculations of Fig.~\ref{fig:pion}, it is interesting to find out that the $A_g$ and $C_g$ form factors have the same hard coeffcient,
\begin{align}
{\cal A}^\pi_g(x_1,y_1,t)={\cal C}^\pi_g(x_1,y_1,t)=\frac{4\pi \alpha_s C_F}{-t}\left(\frac{1}{x_1 \bar x_1}+\frac{1}{y_1 \bar y_1} \right) \ , 
\end{align}
where $C_F=4/3$ and the notation $\bar x=1-x$ is used. A number of interesting features can be found from the above result. First, $A^{\pi}_g$ and $C^{\pi}_g$  GFFs of the pion have the same power counting of $t$. This is different from the nucleon case below, where $C_g^p$ is power suppressed compared to $A_g^p$. Second, they share exactly the same large-$t$ behavior. This is a surprising result. It will be interested to check higher order corrections. In general, we expect this will change.

The hadron GFFs can be derived from the sum rules of the GPDs~\cite{Ji:1996ek}. %For pion GFFs, we have,
%\begin{equation}
%  \int d xH_{q,g}(x,\xi,t)=A_{q,g}(t)+\xi^2 C_{q,g}(t) \ .
%\end{equation}
%where $H_{q,g}$ represent the quark/gluon GPDs. 
%Therefore, we can also calculated the pion GFFs from the quark/gluon GPDs at large $(-t)$. 
The quark GPDs at large momentum transfer have been calculated in Ref.~\cite{Hoodbhoy:2003uu}. We can follow the same procedure to compute the gluon GPD of pion at large momentum, and we find that it leads to the same result for the gluon GFFs as above. This provides an important cross check for our derivations.

In addition, we can derive the quark GFFs for pion from the quark GPD results from Ref.~\cite{Hoodbhoy:2003uu}. In terms of the same factorization formula, we obtain the hard coefficients for $A_q$ and $C_q$ as,
\begin{align}
&{\cal A}_q^\pi(x_1,y_1,t)=\frac{4 \pi \alpha_s C_F}{-t}  \frac{x_1+y_1+1}{\bar x_1 \bar y_1},
\notag \\&
{\cal C}_q^\pi(x_1,y_1,t)=\frac{4 \pi \alpha_s C_F}{-t} 
\frac{x_1+y_1-3}{\bar x_1 \bar y_1}\ .
\label{GFF:pion}
\end{align}
It is interesting to note that, different from the gluon case, $A$ and $C$ form factors are not the same for the quark. However, they have the same power behavior. This is different from the power counting analysis derived in Ref.~\cite{Tanaka:2018wea}. %The physics implication of this result deserves further investigations.
We can also apply the traceless feature of Eq.~(\ref{EMT:pion}) at this order to derive $\overline{C}$ form factors: $\overline{\cal C}_g^\pi=-\overline{\cal C}_{u+\bar d}^\pi=\frac{t}{4m^2}{\cal C}_g^\pi$. The cancellation between the quarks and gluons is expected because of the EMT conservation. Similarly, we find that the $\langle P'|F^2|P\rangle$ form factor of pion does not have power behavior, i.e., it becomes a constant modulo logarithmic dependence from $\alpha_s$ at large $(-t)$. 

Physically, the ${ C}^\pi(t)$ characterizes the mechanical properties such as pressure distribution and shear forces inside the pion system~\cite{Polyakov:2018zvc}. It also determines the mechanical radius of the hadron~\cite{Polyakov:2018zvc,Kumano:2017lhr}. The above results provide important perspectives on these interpretations. 

Another important point from our results is that $C^\pi_g(t)$ is positive at large $(-t)$, whereas there is a strong argument that $C_g^\pi$ is negative at low $(-t)$~\cite{Polyakov:2018zvc} and a recent lattice calculation also confirms that~\cite{Shanahan:2018pib}. That means that $C_g(t)$ will change sign at higher $(-t)$. We hope future lattice simulation can extend to higher momentum transfer to test this prediction. 

\section{Gravitational Form Factor for Nucleon}

Now we turn to investigate the proton cases. Due to its spin, the calculations are more involved. To extract the GFFs, one needs to evaluate the EMT matrix elements for different nucleon helicity configurations. The $A_g(t)$ form factor can be obtained with the helicity-conserved matrix element, whereas $C_g(t)$ requires the helicity-flipped matrix element. Again, we can follow a power counting analysis~\cite{Brodsky:1973kr, Matveev:1973ra,Ji:2003fw} to determine the power behaviors at large $(-t)$. For example, similar to $F_1$ form factor, the $A_g$ form factor scales as $1/(-t)^2$. On the other hand, because of helicity-flip, $C_g$ form factor will scale as $1/(-t)^3$. The detailed calculations below will confirm these power counting analysis. 

 \begin{figure}[tpb]

\includegraphics[width=0.35\columnwidth]{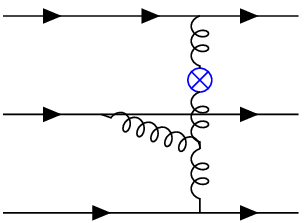}
\includegraphics[width=0.35\columnwidth]{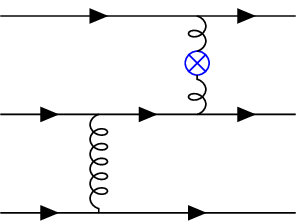}
 \caption{Representative diagrams of two classes that contribute to the gluon GFFs of the proton at the large $-t$ limit. The cross symbol in the middle of diagrams represents the gluonic energy-momentum tensor operator. The three quark lines denote the leading light-cone wave function configuration for the proton state.}
     \label{fig:proton}
 \end{figure}
 
First, we deal with the $A_g(t)$ GFF for the proton. Since it is associated with the proton heilicity-conserved matrix, the procedure toward the factorization will be the same as that for the pion case,
\begin{align}
A_g( t) =&\int [d x][ dy]\Phi_3^*(y_1,y_2,y_3)
  \Phi_3(x_1,x_2,x_3)\nonumber\\
  &\times \  {\cal A}_g(\{x\} ,\{y\}) \ ,
 \end{align}
where $\{x\}=(x_1,x_2,x_3 )$, $[d x]= d x_1 d x_2 d x_3\delta(1-x_1-x_2-x_3)$, and $ \Phi_3(x_i)$ is the twist-three light-cone amplitude of the proton~\cite{Braun:1999te}.   

\begin{figure*}[!htpb]
    \centering 
    
    \includegraphics[width=1.4\columnwidth]{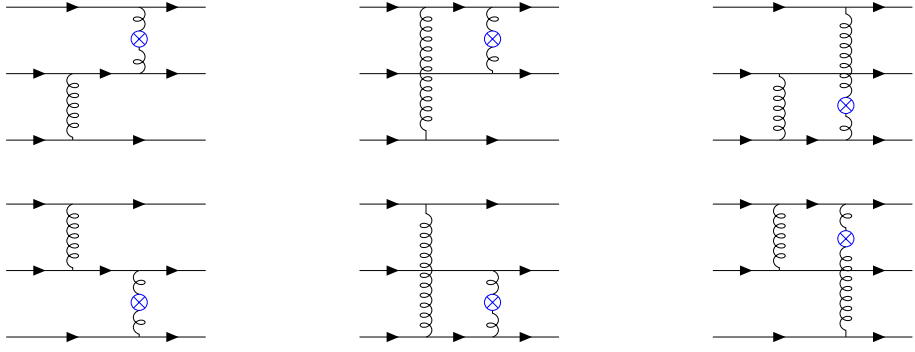}

\caption{Perturbative diagrams that contributes the hard parts of gluon GFFs. Mirrored graphs are implied.}    
\label{fig:proton2}
\end{figure*}

In the calculations, we need to contract the gluonic EMT operator to the three quark light-cone wave function configurations for the initial and final state nucleons. Because of three-gluon vertex in QCD, we have two different classes of diagrams that contribute to the hard part, which are shown in Fig.~\ref{fig:proton}. However, at this order, because of anti-symmetric color structure associated with leading-twist distribution amplitudes in the nucleon states, the diagram in the right panel vanishes. Therefore, we only need to consider the left panel diagram in the perturbative calculations. In this class of diagrams, the local EMT operator is attached to a quark lines by two gluons and another gluon is exchanged separately between two quarks lines. In total, we have 12 diagrams, which are shown in Fig.~\ref{fig:proton2}.

For the $A_g$ form factor, it follows that for the $F_1$ form factor and the contributions from from Fig.~\ref{fig:proton2} can be written as, 
 \begin{align}
  {\cal A }_g(\{x\} ,\{y\}) =2{\cal A}+{\cal A}'.
 \end{align}
where ${\cal A}'$ is obtained from ${\cal A}$ by interchanging $y_1$ and $y_3$. The expression of ${\cal A}$ can be summarized in the following compact form,
\begin{align} 
{ \cal  A}=&\frac{ 4\pi^2\alpha_s^2  C_B^2}{3t^2  }
\Bigl(
  I_{13} + I_{12}+ I_{31}
+I_{32} 
\Bigr),
\end{align}
where $C_B=2/3$ is the color factor. The functions $I_{ij}$ is defined by
\begin{align}
&I_{i j}=\frac{x_i+y_i}{\bar x_i \bar y_i x_i  x_j y_i  y_j } \ .
\end{align}
It has been suggested that the power behavior of the electromagnetic form factors at large $(-t)$ can be related to the power behavior of parton distributions at large $x$~\cite{Drell:1969km,West:1970av}. However, this relation seems break down for the gluonic GFF of nucleon. We know that gluon distribution is $(1-x)$ suppressed respect to the quark distribution~\cite{Brodsky:1994kg}. However, their GFFs have the same power behavior at large $(-t)$, where the quark GFF can be obtained from the GPD calculations in Ref.~\cite{Hoodbhoy:2003uu} (see also the power counting analysis in Ref.~\cite{Tanaka:2018wea}).

Calculation of $C_g$ is much more complicated. This is because it can only be extracted from the helicity flipped matrix element of gluon EMT $\langle P_\uparrow |T_g^{\mu\nu}|P_\downarrow \rangle $ and the final result depends on the higher-twist distribution amplitudes of nucleon. As we mentioned before, it is the quark OAM that generate the proton helicity flip and determine the large momentum transfer behavior of these GFFs.  To include the content of the OAM in the analysis, we follow the strategy and technology in Ref.~\cite{Belitsky:2002kj}. First, we need the  three-quark light-cone Fock expansion  of the proton state \cite{Ji:2002xn}, where the components are denoted with orbital angular momentum $l_z$, e.g. $|P_\downarrow \rangle_{l_z=1}\sim \int  (k^x_1+i k^y_1) \psi_3+ (k^x_3+i k^y_3) \psi_4$, where the factors $(k^x_i+i k^y_i)$ beside the light-cone wave function are the manifestations of the quark OAM. Since the helicities of the up and down quarks are approximately conserved in the high energy scattering, the quark OAM in the intial and final states must differ by one unit. Therefore, the leading contributions will come from the following two matrix elements, 
 $
 {}_{l_z=0}\langle P_\uparrow |(T_g^{\mu\nu})|P_\downarrow \rangle_{l_z=1},
 {}_{l_z=-1}\langle P_\uparrow (T_g^{\mu\nu})|P_\downarrow \rangle_{l_z=0}.
$
 To evaluate this two amplitude, we work in the Breit frame where the initial and finial proton are anti-collinear. In this frame, the partonic quarks have the the longitudinal momenta $x_i P$ and the transverse momentum $\bs k_i$. They emit from the proton and participate in the hard interaction with the gluon EMT operator. Endured with the hard gluon exchanges, these quarks recoil and thus produce the large momentum transfer. Finally, they  obtain the momenta $y_i P'+\bs k'_i$ and recombine into the proton. The collinearity ensures the transverse momenta of the partons is order $\Lambda_{\text{QCD}}$. However, we can not naively ignore the transverse momentum for the leading power. Since the quark OAM  act like $(k^x_i\pm i k^y_i)$ inside the phase space integral, this content of transverse momentum in the hard part  will be picked up by these factors. For that, we should perform the internal transverse momentum expansion on the hard part in the limit of large $-t$. Then only linear terms of quark transverse momentum in the hard part contribute. Therefore, the leading hard part  must have a structure like $\bs k_i {\cal C}(x_1, x_2 ,x_3,y_1, y_2 , y_3 ,t)$. Ultimately, the dependence of $\bs k_i$ will be absorbed in the twist-four amplitude of the proton, e.g. $\Psi_4 \sim \int d^2 \bs k\  \bs k_2 \cdot \{\bs k_1 \psi _3+\bs k_2\psi_4 \}$ .

With the above analysis, we carry out a detailed derivation for all the diagrams of Fig.~\ref{fig:proton2} and $C_g(t)$ can be factorized into,
\begin{align}
&C_g(t) =  \int [d x][d y] \left\{ x_3 \Phi_4(x_1,x_2,x_3)  {\cal C}_{\Phi g}(\{x\},\{y\})\right.\nonumber\\
&
\left.+x_1\Psi_4(x_2,x_1,x_3) {\cal C}_{\Psi g}(\{x\},\{y\})
\right\}\  \Phi_3(y_1,y_2,y_3) \ ,
\end{align} 
where $\Psi_4$ and $\Phi_4$ are the twist-four distribution amplitude of the proton \cite{Braun:2000kw}. ${\cal C}_g$ can be written as,
\begin{align}
{\cal C}_{ g}=2 {\cal C} +{\cal C}',
\end{align}
where ${\cal C}'$ is obtained from ${\cal C}$ by interchanging $y_1$ and $y_3$.  From the detailed calculations of the diagrams in Fig .\ref{fig:proton2}, we obtain
\begin{align}
{\cal C}_{\Psi }(\{x\},\{y\})={\cal H}(\{x\},\{y\}), \quad {\cal C}_{\Phi}={\cal C}_{\Psi }(1\leftrightarrow 3),
\end{align} 
where
  \begin{align}
&{\cal H}(\{x\},\{y\})=\frac{ C_B^2  M^2 }{24(-t)^3} (4\pi \alpha_s)^2  \nonumber\\
&~\times \bigg[
x_3 K_1 
 \left( x_1
   \bar{x}_1+y_1 y_2 - 2 y_3 \bar{x}_1\right)
+
\bar x_3 \tilde K_1 
 \left(x_3 \bar{x}_1+y_3 \bar{y}_3\right)  
\notag \\& ~
+
x_3 (\tilde K_2- K_2) 
\left (x_2 \bar{x}_2 - y_2 \bar{y}_2
\right)
-
K_3
 \bigl(2 \bar{x}_1+y_1\bigr )
\notag \\& ~
+x_3 (K_4 + K_5)
\bigl (x_1-2 \bar{y}_1\bigr)
+
(\tilde K_4+\tilde K_5)
 ( x_3 \bar{x}_3+ y_3 \bar{y}_3  )
 \bigg].
 \end{align}
The functions $K_{i}$ are defined as
\begin{align}
&K_1=\frac{1}{x_1 x_3^2 y_1 y_3^2 \bar{x}_1^2 \bar{y}_1},
\quad
K_2=\frac{1}{x_1 x_2 x_3^2 y_2 y_3^2 \bar{x}_2 \bar{y}_2},\nonumber\\
%\quad
&K_3=\frac{1}{x_1 x_2 y_1 y_2 \bar{x}_1^2 \bar{y}_1}
,
\quad
K_4=\frac{1}{x_1 x_3^2 y_1 y_3 \bar{x}_1 \bar{y}_1^2},
\notag \\
&K_5=\frac{1}
{x_1 x_2 x_3 y_1 y_2 \bar{x}_1 \bar{y}_1^2}, \quad  \tilde K_i=K_i(1\leftrightarrow3).
\end{align}
Comparing the above to the $A_g$ results, we find two important features. First, we confirm the power counting analysis, $C_g$ form factor is suppressed by $1/(-t)$ at large momentum transfer. Second, because of the hard coefficients contain additional factor in the denominator depending on $x_i$ and $y_i$, there will be an end-point singularity in the $C_g$ form factor. We can follow the arguments presented in Ref.~\cite{Belitsky:2002kj} for the Pauli form factor $F_2$ and derive that these end-point singularities will lead to a logarithmic enhancement at large momentum transfer. In the sense, the large $t$ behavior for $C_g(t)$ will be $\ln^2(-t/\Lambda^2)/(-t)^3$ where $\Lambda$ represents a low momentum scale to regulate the end-point singularity in the above integral. The phenomenological importance of these logarithms have been shown for the $F_2$ form factor~\cite{Belitsky:2002kj} and we expect the same for the $C_g^p$ form factor.

It is straightforward to extend the above procedure to all other GFFs and we find that the $B_g$ scales as $\ln^2(-t/\Lambda^2)/(-t)^3$, the same as $C_g$ above, whereas $\overline{C}_g$ scales as $\ln^2(-t/\Lambda^2)/(-t)^2$. Similarly, $\langle P'|F^2|P\rangle $ for the proton scales the same as $\overline C_g$. We emphasize that all these form factors, $B_g$, $C_g$ and $\langle P'|F^2|P\rangle $ come from the helicity-flip amplitude. The power behavior difference between $B_g$ and the latter two is purely due to their parameterization in the form factor definition~\cite{Ji:1996ek,Ji:1996nm}. We will present detailed results for them in a future publication.

\section{Conclusion}

In summary, we have carried out a perturbative analysis of the gluon gravitation form factors for pion and nucleon. The leading order contributions predict that the $C_g^\pi$ form factor is the same as that of $A_g^\pi$ and they both scale as $1/(-t)$ at large momentum. For the nucleon, the $C_g^p$ is power suppressed as compared to the $A_g^p$. Because of the end point singularity, the $C_g^p$ form factor has an additional logarithmic contribution. These results will have profound implications for the phenomenological studies of these form factors and their interpretations as pressure distributions inside hadrons.

Meanwhile, as we mentioned in Introduction, the helicity-conserved quark GPD $H_q$ at large momentum transfer has been calculated in Ref.~\cite{Hoodbhoy:2003uu}. Applying our method in this paper, it will be straightforward to compute all other quark GPDs and the gluon GPDs at large $t$. These results will provide important guidance for future measurements at the EIC~\cite{Accardi:2012qut,Boer:2011fh}, where GPDs and GFFs are among the most important topics to reveal the proton tomography and mass decomposition.

Theoretically, it will be important to investigate further the end-point singularity associated with $C_g^p$ form factor when the quark lines become soft. A rigorous framework needs to be developed where one can factorize and resum these soft parton contributions in the exclusive processes, following, e.g., recent progresses in dealing with the end-point singularity in $H\to \gamma\gamma$ process~\cite{Liu:2019oav,Liu:2020wbn}. We will come back to this issue in a future publication.

{\bf Acknowledgments:} We thank Xiangdong Ji, Maxim Polyakov, Peter Schweitzer, Phiala Shanahan for comments and suggestions. This material is based upon work supported by the U.S. Department of Energy, Office of Science, Office of Nuclear Physics, under contract numbers DE-AC02-05CH11231. J.P. and X.B. are supported by National Natural Science Foundation of P.R. China(No.12075299,11821505, 11935017)  and by the Strategic Priority Research Program of Chinese Academy of Sciences, Grant No. XDB34000000. X.B acknowledges the scholarship provided by the University of Chinese Academy of Sciences for the joint Ph.D. training.

\bibliographystyle{apsrev4-1}
\bibliography{references}

\end{document}